\documentclass[%
 reprint,
 amsmath,amssymb,
 aps,
 pra,
]{revtex4-2}

\usepackage{hyperref}
\usepackage{qcircuit}
\usepackage{tikz}
\usepackage{physics}
\usepackage{algorithm}      
\usepackage{algorithmic} 
\usepackage{amsmath} 

\begin{document}

\title{Enhanced security in Quantum Token protocols using Hybrid Spin-Photon Interfaces}

\author{Durga Bhaktavatsala Rao Dasari}
\email{d.dasari@pi3.uni-stuttgart.de}
\author{Yang Wang}
\author{J\"{o}rg Wrachtrup}
\affiliation{%
 3. Physikalisches Institut, ZAQuant, University of Stuttgart, 70569 Stuttgart GERMANY
}%

\date{\today}

\begin{abstract}
Quantum token protocols enable unforgeable quantum tokens promising unconditional security beyond classical cryptographic assumptions. We show here that the three stages of the Quantum token protocols involving the preparation, storage and verification can be made more secure when involving spin-photon interfaces that leverage high fidelity hybrid tripartite (spin-photon-spin) entanglement, Bell state measurements and highly coherent spin quantum memories. Further we describe the physical implementation of various stages of the protocol using the hybrid electron and nuclear spins in diamond interfaced with time-bin photons.

\end{abstract}

\maketitle

\section{Introduction}

Quantum token protocols have emerged as a promising paradigm in quantum cryptography, enabling the creation of unforgeable quantum tokens that leverage fundamental principles of quantum mechanics such as the no‑cloning theorem and measurement disturbance \cite{Wiesner1983_ConjugateCoding}. Subsequent theoretical developments established rigorous security models and explicit constructions of quantum money and token schemes \cite{AaronsonChristiano2012_HiddenSubspaces, Lutomirski2009_BreakingMakingQM}, highlighting their advantages over classical cryptographic tokens that rely on computational assumptions. Unlike classical tokens, quantum tokens offer unconditional security guaranteed by the laws of physics, making them highly relevant for applications in secure authentication, digital cash, and copy‑protection. The verification procedures in these protocols often exploit properties of quantum states that enable public verification without revealing the token’s secret information, thus allowing for transferability and privacy. While conventional quantum token schemes involve quantum memories and long‑distance quantum communication, their counterparts in quantum optical money have been experimentally implemented using heralded single photon sources and classical verification steps \cite{Bartkiewicz2017_QuantumOpticalMoney, Jiang2024}.
On the other hand, quantum tokens differ fundamentally from quantum digital signatures and identity authentication protocols \cite{QDS}. While digital signatures ensure the integrity and authenticity of a message, and identity schemes verify a user’s possession of a secret, quantum tokens function as transferable, unforgeable value objects that can be securely stored and verified at a later time. Here we propose a QT-protocol that uniquely leverages entangled spin-photon interfaces to issue and teleport tokens securely, a functionality that neither standard quantum signatures nor identity authentication schemes provide.

Practical implementations, particularly through the integration of hybrid quantum systems that combine spins, photons, and mechanical resonators \cite{Kurizki2015_HybridQuantumSystems} can be advantageous for the QToken schemes. Such hybrid platforms leverage the long coherence times of spin qubits alongside the communication advantages of photonic qubits and the strong coupling facilitated by phononic modes, addressing critical challenges in token storage, transmission, and scalability. Moreover, contemporary protocols increasingly incorporate classical cryptographic primitives, including blind signatures and zero‑knowledge proofs, to enhance user privacy and robustness in adversarial settings \cite{Ribeiro2015_QuantumBlindSignature, Shi2015_UnlinkableBlindSig}. Experimentally, milestones in high‑fidelity quantum memories \cite{England2012_RAMANMemory}, entanglement distribution \cite{Hensen2015}, and single‑photon generation \cite{Senellart2017} have laid the groundwork for realizing quantum tokens in realistic networked environments. Collectively, these developments position quantum token protocols as foundational components for emerging quantum‑secure infrastructures such as the quantum internet, with significant implications for financial security, identity authentication, and anti‑counterfeiting technologies.

\begin{figure*}[htbp]
%\centering
\includegraphics[width=1.0\textwidth]{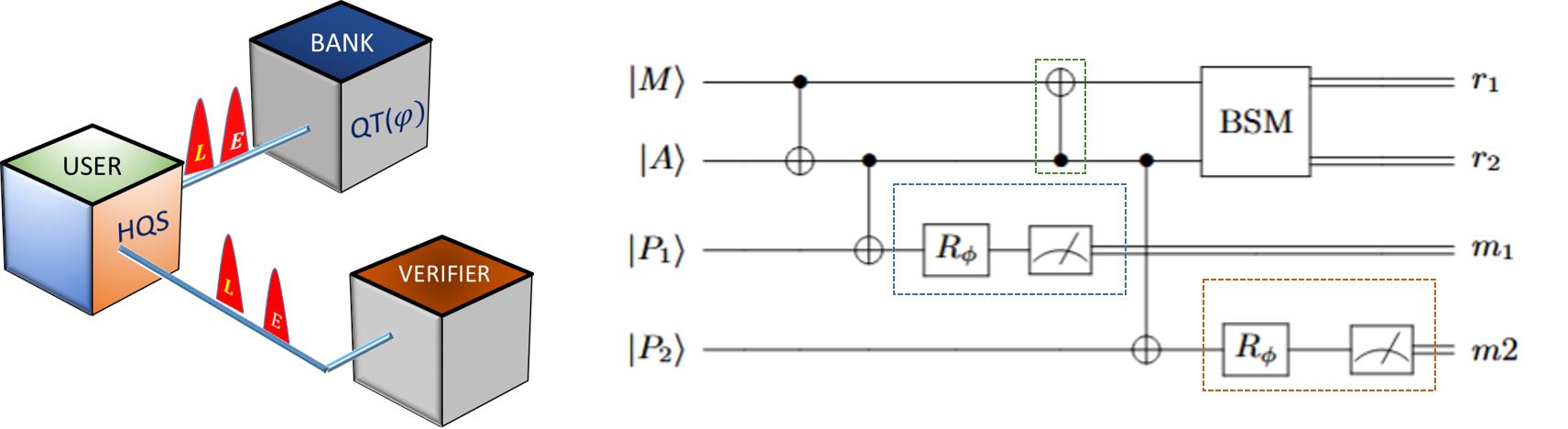}
\caption{(Left) Schematic displaying the three entitites of the QT protocol, viz., the User, Bank and the Verifier.  The user communicates both with the bank and the verifier by sending time-bin photons that are entangled to the ancillay spins of his Hybrid Quantum node. (Right) The Quantum circuit describing all the seven stages of the protocol is shown. The QT issuance storage and verification parts are highlighted in the circuit. From the classical measurement output data one establishes the strong verification conditions i.e.,  $m2=m1 \oplus (r1 \oplus r2)$. }
\label{fig:my_circuit}
\end{figure*}

Nitrogen‑vacancy (NV) centres in diamond have emerged as a leading solid‑state platform for realizing spin‑photon entanglement and quantum memories due to their exceptional coherence properties and optical addressability at room temperature \cite{Doherty2013, vanDam2012_Coherence}. The NV centre’s electron spin state can be initialized, manipulated, and read out optically via spin‑selective fluorescence transitions, enabling high‑fidelity spin‑photon interfaces \cite{Togan2010, Awschalom2018_QuantumInterfacedSpins}. Spin‑photon entanglement protocols exploit the coupling between the NV electron spin and emitted photons, which can be generated in well‑defined polarization and frequency modes, thus serving as flying qubits for quantum communication \cite{Togan2010, Bernien2013}. The ability to coherently transfer quantum information between stationary electron spins and photonic qubits is critical for quantum network applications, including entanglement distribution, quantum repeaters, and secure quantum token protocols. Furthermore, NV optical transitions can be selectively addressed even at cryogenic temperatures, facilitating coherent control of the electron spin‑photon system with minimal decoherence \cite{vanDam2018}.

In addition to electron spin control, the hyperfine interactions between the NV center electron spin and nearby nuclear spins, primarily $^{13}$C nuclei, provide a robust resource for quantum memory and error correction \cite{Childress2006_CoherentEN, Zhang2025_NuclearDephasingNV}. Nuclear spins exhibit significantly longer coherence times than electron spins, often extending into seconds at low temperature, making them ideal candidates for quantum storage. Controlled electron‑nuclear spin interactions allow for the transfer and retrieval of quantum states between the fast‑manipulated electron spin and the long‑lived nuclear spin registers, thereby extending the effective coherence time of the system \cite{Zhang2025_NuclearDephasingNV}. This electron‑nuclear spin coupling also enables multi‑qubit quantum registers localized at a single NV center, facilitating the implementation of small‑scale quantum processors and error‑resilient quantum memories \cite{Abobeih2018_OneSecondCoherence}. Collectively, the exceptional spin coherence, optical interface, and controllable electron‑nuclear interactions position NV centers as versatile building blocks for hybrid quantum networks and advanced quantum information processing protocols, including quantum tokens that require reliable storage, manipulation, and distribution of quantum states. Further the ability to entangle time-bin photons with both the electron and nuclear spins \cite{Tchebotareva2019spinPhotonTimeBin, javid}can become a resource for such protocols that we show below.

\section{Protocol}
%We consider a classical-quantum hybrid token protocol involving three principal entities: the \textit{User} who possesses the quantum token, given by the \textit{Bank} which acts as the trusted issuer, and the \textit{Verifier} responsible for authenticating the token of the user at a later stage. The protocol aims to securely issue, transfer, store, and verify these quantum tokens when required. All these stages are to be secured using fundamental quantum principles, while leveraging classical cryptographic methods to ensure privacy and robustness.

%Our model the user is a long-lived nuclear spin (Quatum Memory), and the role of bank issuing the token is taken by a photon detector/projector, and an electron spin that mediates (interface) the connection between the user and the bank. The verifiers role is taken by a different photon detector/projector. We detail below the steps involved in the protocol:

We consider a classical-quantum hybrid token protocol involving three principal entities: the \textit{User} who possesses the quantum token, given by the \textit{Bank} which acts as the trusted issuer, and the \textit{Verifier} responsible for authenticating the token of the user at a later stage. The protocol aims to securely issue, transfer, store, and verify these quantum tokens when required. All these stages are to be secured using fundamental quantum principles, while leveraging classical cryptographic methods to ensure privacy and robustness.

In our model the user can represent a long-lived quantum memory (M), and the role of the bank issuing the token is taken by a photon detector/projector ($P_1$), and an ancilla spin (A) that mediates (interface) the connection between the user and the bank and the verifier (see Fig. 1). The verifiers role is taken by a different photon detector/projector ($P_2$). We detail below the steps involved in the protocol:

\vspace{0.2cm}
  \textbf{Step 1: A-M Entanglement (A-M)}
  Prepare entangled state between A and M spins:
\begin{equation}
    |\psi_{A,M}\rangle = \frac{1}{\sqrt{2}} \left( |0\rangle_A |\uparrow\rangle_M + |-1\rangle_A |\downarrow\rangle_M \right)
\end{equation}
using standard quantum gates (e.g., CNOT and Hadamard).

\textbf{Step 2: A-Photon ($A-P_1$) Entanglement (Time-bin Encoding)}
The user contacts the bank by sending a photon that is entangled to its ancillary qubit $A$. This entanglement between the ancilla and the photon $P_1$ is established through time-bin encoding. As the ancilla is already entangled to the memory qubit, we obtain the combined tripartite state 
\begin{equation}
|\Psi\rangle = \frac{1}{\sqrt{2}} \left( |0\rangle_A  |\uparrow\rangle_M |E\rangle_{p_1} + |-1\rangle_A  |\downarrow\rangle_M |L\rangle_{p_1} \right)
\end{equation}
where $|E\rangle, |L\rangle$ are early and late time-bin photon states.

\textbf{Step 3: Project $P_1$ in target basis:}
The role of the bank is to take the user sent photon, rotate and measure $P_1$, by projecting it onto one of the basis states, $|\pm_\phi\rangle= \frac{1}{\sqrt{2}} \left( |E\rangle \pm e^{i\phi} |L\rangle \right)$ (see Fig. 2a). The measurement yields the classical bit $M_1$, and the tripartite state above collapses to the state,  
\begin{equation}
|\psi_{A,M}^{\pm}\rangle = \frac{1}{\sqrt{2}} \left( |0\rangle_A |\uparrow\rangle_M \pm e^{-i\phi} |-1\rangle_A |\downarrow\rangle_M \right)
\end{equation}

\textbf{Step 4: Storage of the Quantum Token}
 The bank issued Quantum Token is stored in the memory qubit by disentangling the $A-M$ qubits 
\begin{equation}
|\psi_M\rangle = \frac{1}{\sqrt{2}} \left( |\uparrow\rangle_M  \pm e^{-i\phi} |\downarrow\rangle_M \right)
\end{equation}

  \textbf{Step 5: A-Photon Entanglement}
After a certain storage time $T_S$, the user again contacts the verifier by sending a photon $P_2$ that is entangled to its ancillary spin. Using a similar time-bin encoding method, the new entangled state between the users qubit $A$  and the photon $P_2$ is
\begin{equation}
|\psi_{A,p_2}\rangle = \frac{1}{\sqrt{2}} \left( |0\rangle_A |E\rangle_{p_2} + |-1\rangle_A |L\rangle_{p_2} \right)
\end{equation}

  \textbf{Step 6: A-M Bell State Measurement (BSM)}
  We will now teleport the stored token qubit to the verifier by performing a BSM on the user qubits $A, M$. This measuremnet yields two classical bits $(r_1, r_2)$.

  \textbf{Step 7: Final Measurement of $P_2$}
The verifier with measure the photon $P_2$ in the secret basis set by the bank $\{ |+_\phi\rangle, |-_\phi\rangle \}$ basis, and obtain the result $M_2$.

\subsubsection*{Security and Error Analysis}

The quantum token protocol is subject to several operational errors: (i) imperfect A--M preparation, (ii) random phase noise in the two A--P entanglement steps, (iii) memory dephasing during storage time $T_S$, (iv) Bell-state–measurement imperfections $F_{\rm BSM}$, and (v) possible basis mismatch $\Delta \phi$.  
We model these effects as follows. Let the initial A--M state deviate from a perfect Bell state,
\[
|\psi_{A,M}\rangle = \sqrt{\alpha}\,|0\uparrow\rangle + \sqrt{1-\alpha}\,|-1\downarrow\rangle,
\]
and let the first ancilla--photon entanglement introduce an unknown phase $\theta_1$. After projecting the photon and disentangling the ancilla, the stored memory token becomes
\[
|\psi_M\rangle = \sqrt{\alpha}\,|\uparrow\rangle_M + \sqrt{1-\alpha}\, e^{i (\theta_1-\phi_1)}\,|\downarrow\rangle_M.
\]

During verification, a second photon entanglement can introduce another phase $\theta_2$, giving $\Delta \theta = \theta_1-\theta_2$. Combined with the measurement basis mismatch $\Delta \phi = \phi_1-\phi_2$, the verification fidelity (assuming ideal BSM) is
\[
F_\mathrm{verif}(\alpha, \Delta \theta) = \frac{1}{2}\Big[ 1 + 2 \sqrt{\alpha(1-\alpha)} \cos(\Delta \theta - \Delta \phi) \Big].
\]

Averaging over uniform $\alpha \in [0,1]$ and Gaussian phase noise $\Delta \theta \sim \mathcal{N}(0,\sigma_\theta^2)$ gives a device-independent metric including operational errors. Incorporating memory decoherence $e^{-T_S/T_M}$, we obtain
\[
\langle F_\mathrm{verif} \rangle = \frac{1}{2} \Big[ 1 + \frac{\pi}{4} \, e^{-T_S/T_M} \, e^{-\sigma_\theta^2/2} \cos(\Delta \phi) \Big].
\]

This averaged verification fidelity directly defines the security metrics. For an honest user, $\langle F_\mathrm{verif} \rangle$ quantifies completeness—the probability that a valid token is accepted. A dishonest adversary without access to the memory can only guess, giving a maximum forgery probability $F_\mathrm{forge}^{(1)} \le 1/2$. The soundness error, i.e., the probability that a forged token is accepted, is then
\[
\varepsilon_\mathrm{sound} = \langle F_\mathrm{verif} \rangle - F_\mathrm{forge},
\]
which remains positive as long as coherence is preserved. Phase noise $\sigma_\theta$ and basis mismatch $\Delta \phi$ appear explicitly in $\langle F_\mathrm{verif} \rangle$, providing a quantitative measure of security degradation under realistic imperfections. The strong verification condition
\[
m_2 = m_1 \oplus (r_1 \oplus r_2)
\]
follows from teleportation-induced Pauli corrections, ensuring that only a valid token satisfying the correlations between the bank’s measurement, the BSM outcomes, and the verifier’s measurement can be accepted; any adversary without the memory and BSM succeeds with probability at most $1/2$.

\section{Physical Implementation}
We begin by initializing both the NV electron spin (ancilla) and the $^{14}$N nuclear spin (memory) into the state $|0\rangle_a |\uparrow\rangle_m$. To generate entanglement between the two, we first apply an RF $\pi/2$ pulse to the nuclear spin, preparing it in an equal superposition state: $\frac{1}{\sqrt{2}}[|\uparrow\rangle_m + |\downarrow\rangle_m]$. Keeping the electron spin in $|0\rangle_a$, we then apply a conditional $\pi$-rotation (a CNOT gate) on the electron spin, controlled by the nuclear spin. This produces the entangled state
\begin{equation}
|\psi_{a,m}\rangle = \frac{1}{\sqrt{2}} \left( |0\rangle_a |\uparrow\rangle_m + |-1\rangle_a |\downarrow\rangle_m \right).
\end{equation}

Following this, we perform spin-photon entanglement using a resonant, time-bin encoded excitation scheme applied to the NV electron spin. Optical transitions $E_{1,2}$ and $E_y$ are used for spin initialization and coherent excitation, respectively. {Starting from the above ancillary-memory entangled state an optical laser excitation pulse along the transition $E_y$ leads to an early photon emission into the zero-phonon line, conditioned to NV center spin state ${m_s = 0}$. Afterward, a second optical excitation pulse is applied following the MW pulse ($|0\rangle_a \leftrightarrow |-1\rangle_a$), which emits a \textit{late} photon if the electron spin is in $|-1\rangle_a$.}  This sequence results in the tripartite entangled state
\begin{equation}
|\Psi\rangle = \frac{1}{\sqrt{2}} \left( |0\rangle_a |\uparrow\rangle_m |E\rangle_{p_1} + |-1\rangle_a |\downarrow\rangle_m |L\rangle_{p_1} \right),
\end{equation}
where $|E\rangle$ and $|L\rangle$ denote the early and late time-bin photon states, respectively.

\begin{figure}[ht]
\includegraphics[width=0.5\textwidth]{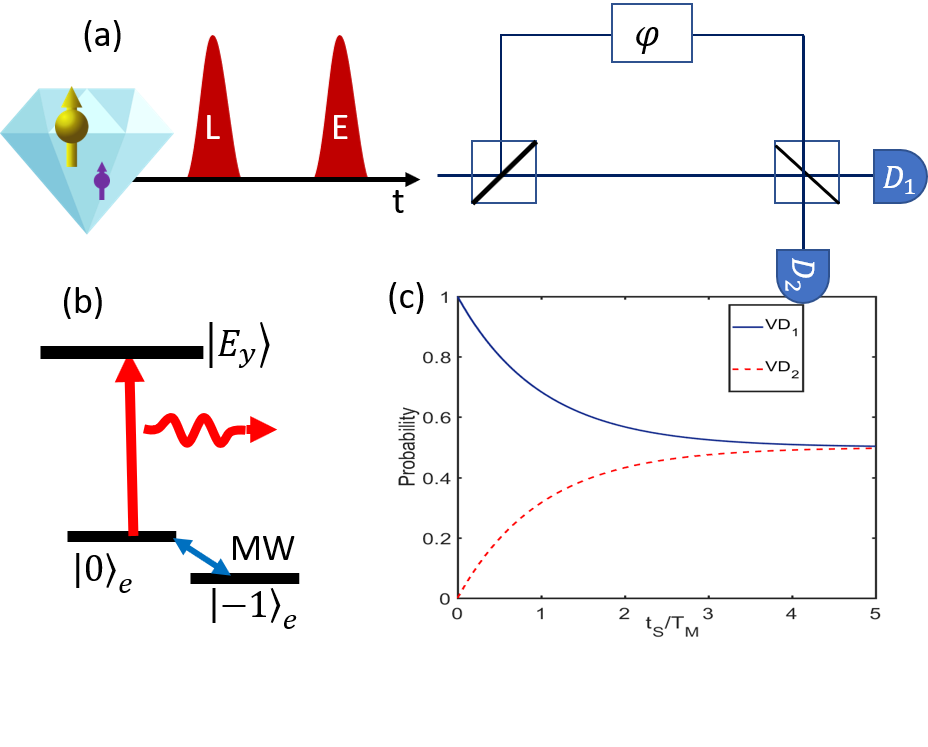}
\caption{(a) Schematic of the physical system implementing the QT protocol. The NV center electron spin in diamond (yellow arrow) coupled to a nearby nuclear spin (purple arrow) emitting time-bin photons that are respectively phase-shifted ($\varphi$) and detected at $D_{1(2)}$. Both the bank and the verifier use this setup for issuing and verifying their Quantum Tokens. (b) The energy-level diagram showing the states and control invovled in establishing the spin-photon interface. (c) The probiability of verifying the user token by clicks in the verifiers detectors, $VD_{1(2)}$ is plotted as a fucntion of the storage time $T_S$ measured in units of the memory life-time $T_M$.  }
\end{figure}

The spin-photon entangled state generated by the NV center is sent to the bank, where the photon undergoes further processing. The bank’s setup includes an unbalanced Mach-Zehnder interferometer equipped with a fiber stretcher that imparts a controlled relative phase $\phi$ between the photon’s two spatial modes, as well as two detectors, $D_1$ and $D_2$, for photon measurement.

Starting from the initial spin-photon state,
\begin{equation}
|\Psi_1\rangle = \frac{1}{\sqrt{2}} \left( |0\rangle_a |L\rangle + |-1\rangle_a |E\rangle \right),
\end{equation}
the photon’s early ($|E\rangle$) and late ($|L\rangle$) time-bin components are directed through different paths of the interferometer, corresponding to the long and short arms respectively. The fiber stretcher inside the long arm introduces a phase shift $\phi$ between the $\text{Long}$ and $\text{Short}$ spatial modes, resulting in the state before the second beam splitter,
\begin{equation}
|\Psi_3\rangle = \frac{1}{\sqrt{2}} \left( |0\rangle_a |\text{Short}\rangle + e^{i\phi} |-1\rangle_a |\text{Long}\rangle \right).
\end{equation}

At the second beam splitter, the photon is coherently split into two output paths leading to detectors $D_1$ and $D_2$, yielding the state

\begin{align}
|\Psi_4\rangle = \frac{1}{2} \big(& |D_1\rangle \otimes \left(|0\rangle_a |\uparrow\rangle_m + e^{i\phi} |-1\rangle_a  |\downarrow\rangle_m\right) \nonumber \\
+\, & |D_2\rangle \otimes \left(|0\rangle_a |\uparrow\rangle_m - e^{i\phi} |-1\rangle_a  |\downarrow\rangle_m\right) \big)
\end{align}

A photon detection event in $D_1$ projects the electron-nuclear spin system onto the state
\begin{equation}
|\psi_{a,m}^{+}\rangle = \frac{1}{\sqrt{2}} \left( |0\rangle_a |\uparrow\rangle_m + e^{i\phi} |-1\rangle_a |\downarrow\rangle_m \right),
\end{equation}
while detection in $D_2$ projects it onto
\begin{equation}
|\psi_{a,m}^{-}\rangle = \frac{1}{\sqrt{2}} \left( |0\rangle_a |\uparrow\rangle_m - e^{i\phi} |-1\rangle_a |\downarrow\rangle_m \right).
\end{equation}
Thus, the bank’s interferometric setup with phase control and photon detection heralds a known phase encoding and projects the joint ancilla-memory spin state accordingly.
Following the photon measurement at the bank, the electron (ancilla) and nuclear (memory) spins remain entangled in the state
\begin{equation}
|\psi_{a,m}^{\pm}\rangle = \frac{1}{\sqrt{2}} \left( |0\rangle_a |\uparrow\rangle_m \pm e^{i\phi} |-1\rangle_a |\downarrow\rangle_m \right).
\end{equation}
To store the phase information $\phi$ encoded by the bank into the nuclear memory spin, we apply the same CNOT gate used earlier for entanglement generation, but now to disentangle the spins. This CNOT operation is performed with the nuclear spin as control and the electron spin as target, effectively transferring the relative phase onto the memory spin. After this operation, the electron spin is disentangled, and the nuclear spin is left in the state
\begin{equation}
|\psi_m\rangle = \frac{1}{\sqrt{2}} \left( |\uparrow\rangle_m \pm e^{i\phi} |\downarrow\rangle_m \right),
\end{equation}
thereby storing the bank’s token $\phi$ securely within the memory spin.

For the verification stage, the user will contact the verifier by sending a single photon that is entangled with it ancillary qubit. For this we will start by applying a microwave (MW) $\pi/2$ pulse to prepare the electron spin into a coherent superposition state, $|\psi_0\rangle_a = \frac{1}{\sqrt{2}} \left( |0\rangle_a + |-1\rangle_a \right)$.
Next, a resonant optical excitation pulse on the $E_y$ transition is applied, which conditionally emits an \textit{early} photon if the electron spin is in $|0\rangle_a$. A MW $\pi$ pulse then swaps the electron spin populations, $|0\rangle_a \leftrightarrow |-1\rangle_a$, followed by a second optical excitation pulse on the $E_y$ transition that conditionally emits a \textit{late} photon if the electron spin is in $|-1\rangle_a$.
This sequence results in the entangled electron-photon state:
\begin{equation}
|\psi_{a,p_2}\rangle = \frac{1}{\sqrt{2}} \left( |0\rangle_a |E\rangle_{p_2} + |-1\rangle_a |L\rangle_{p_2} \right),
\end{equation}
where $|E\rangle$ and $|L\rangle$ denote early and late time-bin photon states, and $\phi$ is the controllable relative phase.

As the photon departs from the user node, a Bell State Measurement (BSM) is performed on the combined electron-nuclear spin system. This measurement effectively teleports the nuclear spin state onto the second photon. Measuring this photon in the appropriate basis—matching the original bank encoding—enables user verification. Since nuclear spin control and measurements are relatively slower compared to the ancilla and photon dynamics, we first SWAP the nuclear spin state onto the electron spin. This results in the electron spin being in the state $\frac{1}{\sqrt{2}} \left( |0\rangle_a \pm e^{i\phi} |-1\rangle_a \right).$
Performing time-bin encoding and projecting the electron spin onto the $\ket{\pm}_a$ basis then projects photon $P_2$ into the states
$|E\rangle_{p_2} \pm e^{i\phi} |L\rangle_{p_2}.$
Verification is completed only upon receiving the classical measurement outcomes from the user, ensuring that any unwanted phase-flips (decoherence) during the storage is properly taken into account. In practical applicatoins where the protocol must be repeated $n$ times to achieve a successful Ancilla-Photon entanglement or photon detection events, each repetition represents an independent trial for the adversary to learn about the token. 
Assuming the adversary attempts to forge in each trial, the overall success probability after $n$ repetitions is then
\begin{equation}
F_\mathrm{forge}^{(n)} \lesssim \left( \langle F_\mathrm{verif} \rangle \right)^n,
\end{equation}
showing that the protocol’s effective unforgeability improves exponentially with the number of independent repetitions. 

In conclusion, we have presented a simple practical and robust Quantum Token protocol in which token issuance, storage, and verification are carried out using multipartite entangled states. At no stage of the protocol are quantum information-carrying photons accessible to potential eavesdroppers, thereby guaranteeing unforgeability. Further the measurement results $[m_1, m_2, r_1, r_2]$ can form the classical token for the corresponding quantum token to enhance the security. The importance of a Quantum Hybrid Node—featuring high-fidelity spin-photon and spin-spin interfaces—within a quantum communication network is emphasized as a key component for the implementation of such protocols. This scheme is readily scalable, allowing each user node to store and manage more than 20 Quantum Tokens, which can be verified on demand. Looking ahead, such tokens could also serve as cryptographic keys for performing secure computations on a cloud-based quantum computer

\begin{acknowledgments}
We would like to acknowledge the financial support by the Federal Ministry of Research, Technology and Space (BMFTR) project QECHQS (Grant No. 16KIS1590K).

\end{acknowledgments}

\bibliographystyle{apsrev4-2}
\bibliography{refs}

\end{document}